\documentstyle[aps,prd,amsfonts]{revtex}
\begin{document}
\draft
\title{
The gauge invariance of general relativistic tidal heating 
}

\author{Patricia Purdue} 

\address{
Theoretical Astrophysics, California Institute of Technology,
Pasadena, CA 91125 
}

\date{Sumbitted 29 January 1999; revised 15 June 1999}
\maketitle
\begin{abstract}

When a self-gravitating body (e.g., a neutron star or black hole) interacts
with an external tidal field (e.g., that of a binary companion), the
interaction can do work on the body, changing its mass-energy.
The details of this ``tidal heating" are analyzed using the Landau-Lifshitz
pseudotensor and the local asymptotic rest frame of the body.  It is shown that
the work done on the body is gauge-invariant,
while the body-tidal-field interaction energy contained within the body's
local asymptotic rest frame is gauge dependent.  This is analogous to 
Newtonian theory, where 
the interaction energy is shown to depend on how one localizes 
gravitational energy, 
but the work done on the body is independent of that localization.

\end{abstract}

\narrowtext
\twocolumn

\section{Introduction and Summary}
\label{sec:Introduction}

This is one in a series of papers that develops perturbative
mathematical and physical
tools for studying the interaction of an isolated gravitating body with a
complicated ``external universe'' in the slow-motion limit.  By ``slow-motion
limit'' we mean that the shortest timescale $\tau$ for changes of the body's
multipole moments and/or changes of the universe's tidal gravitational field
is long compared to the body's size $R$:  $R/\tau \ll 1$, where 
we have set the speed of light equal to unity.

In addition to this slow-motion requirement, we also require that the
body be isolated from other objects in the external universe, in the sense
that both the radius of curvature $\cal R$ of the external universe in the
body's vicinity, and the lengthscale $\cal L$ on which the universe's 
curvature changes there, are long compared to
the body's size: $R/{\cal R} \ll 1$ and $R/{\cal L} \ll 1$.

The slow-motion, isolated-body formalism, to which this paper is a technical
addendum, is, in essence, a perturbative expansion in powers of the small
parameters $R/\tau$, $R/{\cal R}$ and $R/{\cal L}$.  For a detailed discussion
of the structure of this expansion and its realm of validity, see Thorne and
Hartle \cite{ThorneHartle}.  As they discuss at length (their Sec.\ I.B), the
slow-motion and isolated-body assumptions make no reference to the internal
gravity of the object under study.  Consequently, the Thorne-Hartle formalism
in general, and the results of this paper in particular, can be
applied even to strongly-gravitating bodies, as long as the source of the
external tidal field is far enough away to allow a ``buffer zone" where 
gravity is weak.  This buffer zone, called the local asymptotic rest frame, 
will be described more fully at the beginning of Sec.\ III.

Two examples of 
isolated, slow-motion bodies are: (i) a neutron
star or black hole in a compact binary system that spirals inward due to
emission of gravitational waves; and (ii) the satellite Io, which travels around
Jupiter in an elliptic orbit and gets heated by Jupiter's tidal 
gravitational field \cite{Peale}.

The series of papers that has been developing 
 the perturbative formalism  
for studying tidal
effects in such slow-motion, isolated bodies is:

1. The book {\it Gravitation} \cite{MTW}, Section 20.6 (written by 
John Wheeler): laid the
conceptual foundations for analyzing the motion of such an isolated
body through the external universe.

2. Thorne and Hartle \cite{ThorneHartle}: formulated the problem of
analyzing the effects of the external universe's tidal fields on such
an isolated body, and conceived and 
 initiated the development of the perturbative 
formalism
for studying the influence of the tidal fields on the body's motion
through the external 
 universe, the precession of its spin axis, and its changes of 
mass-energy. 

3. Thorne \cite{Thornermp}: developed the theory of
multipole moments of the isolated body in the form used by Thorne and
Hartle.

4. Zhang \cite{Zhang34}: developed the theory
of multipole moments for the external universe's tidal gravitational
fields, which underlies the work of Thorne and Hartle.

5. Zhang \cite{Zhang31}: extended the Thorne-Hartle
analysis of motion and precession to include higher order moments than they
considered.

6. Thorne \cite{TidalHeating} and Flanagan \cite{Flanagan}: initiated the 
study of tidally-induced volume changes in the isolated body,
using the above formalism.  Their studies were motivated by numerical
solutions
of Einstein's equations by Wilson, Mathews, and Maronetti 
\cite{WilsonMathews} which seemed
to show each neutron star in a binary being compressed to the point of collapse
by gravitational interaction with its companion.  Thorne and Flanagan found
no such effect of the large magnitude seen in the numerical solutions.  An
important piece of Thorne's analysis came from examining the work done on 
each star by
its companion's tidal field---i.e., an analysis of ``tidal heating.''

Thorne's analysis of tidal heating required dealing with an issue that
Thorne and Hartle had discussed, but avoided confronting:  For an isolated body
with
mass quadrupole moment ${\cal I}_{jk}$, being squeezed by an external tidal
gravitational field ${\cal E}_{jk} \equiv R_{j0k0}$
(with $R_{\alpha\beta\gamma\delta}$ the external Riemann tensor), there
appears
to be an ambiguity in the body's total mass-energy $M$ of order $\delta M
\sim
{\cal I}_{jk} {\cal E}_{jk}$.  (Here and throughout we use locally Cartesian
coordinates in the body's local asymptotic rest frame; cf.\ Thorne and
Hartle \cite{ThorneHartle}.  Because the coordinates are Cartesian, it
makes no difference whether tensor indices are placed up or down.)

One can understand this apparent ambiguity by examining the time-time part
of the spacetime metric in the body's local asymptotic rest frame:
\begin{equation}
g_{00} = -1+{2M\over r} + {3 {\cal I}_{jk}n^jn^k\over r^3} + \ldots -
{{\cal E}_{jk}n^jn^k r^2} + \ldots \;.
\end{equation}
Here $n^j = x^j/r$ is the unit radial vector and $r$ is distance from the
body
in its local asymptotic rest frame.  Among the terms omitted here are those
of quadratic and higher order in the body's mass $M$ and quadrupole moment
${\cal
I}_{jk}$ and the external tidal field ${\cal E}_{jk}$---terms induced by
nonlinearities of the Einstein field equations.  Among these nonlinear
terms is
\begin{equation}
\delta g_{00} \sim { {\cal I}_{jk} {\cal E}_{jk} \over r} \;,
\end{equation}
whose $1/r$ behavior can be deduced by dimensional considerations.  This
term
has the multipolar structure (monopole)$/r$ identical to that of the $2M/r$
term from which one normally reads off the body's total mass, and its numerical
coefficient is ambiguous corresponding to the possibility to move some
arbitrary portion of it into or out of the $2M/r$ term.  Correspondingly,
the body's mass is ambiguous by an amount of order
\begin{equation}
\delta M \sim {\cal I}_{jk} {\cal E}_{jk} \;.
\label{eq:delM}
\end{equation}

In Sec.\ II of this paper we shall see that this ambiguity is not a purely
relativistic phenomenon.  In the Newtonian theory of
gravity, there is also an ambiguity of magnitude (\ref{eq:delM}) in that 
portion of the 
gravitational interaction
energy of the body and external field which is contained within the
body's local asymptotic rest frame.  In Sec.\ III we shall explore
the relativistic version of this ``tidal-quadrupolar" interaction ambiguity 
by computing how the interaction
energy
changes under a relativistic change of gauge (infinitesimal coordinate
transformation).

``Tidal heating" of the isolated body involves injecting into it an amount of
energy that is of just the same magnitude as this ambiguity,
$\Delta M \sim {\cal I}_{jk} {\cal E}_{jk}$.  Does this mean that
tidal heating is an ill-defined, unphysical concept?  Certainly not, as
is attested by photographs of volcanic plumes ejected from Jupiter's
satellite Io (see, e.g., Ref.\ \cite{Science}).  That volcanism was predicted 
by Peale, Cassen, and Reynolds \cite{Peale} before the Voyager
spacecraft discovered it; and their explanation---that Io gets tidally
heated
to high internal temperature by the coupling of its quadrupole
moment to Jupiter's tidal field---is firm (see, e.g., Refs.
\cite{Ross,Ojak}).

In this paper, we use the phrase ``tidal heating" to mean the net work done by
an external tidal field on an isolated body.  This phrase is slightly 
misleading, as
the work done on the body need not necessarily go only into heat.  The 
additional energy might be used to deform the body (i.e., 
raise a tide on it) or it might go into vibrational energy.  If, however, 
timescales for changes of ${\cal E}_{jk}$ and ${\cal I}_{jk}$ are not close
to the body's vibrational periods, then, when averaged over many cycles of
change of ${\cal E}_{jk}$ and/or ${\cal I}_{jk}$, the work will
contribute primarily to heat, as in the case of Io.

In his analysis of binary neutron star systems, Thorne \cite{TidalHeating} 
argued, but did
not prove in detail, that although the tidal-quadrupole interaction energy is
ambiguous, the amount of work done on an isolated body by an external tidal
field (i.e., the amount of tidal heating) is unambiguously given by the
formula\footnote{ Actually, 
expression (\ref{eq:work}) is only the leading order
term in the perturbative expansion of $dW/dt$.  The next term is
$-{2\over3}{\cal B}_{ij} d{\cal S}_{ij}/dt$, where ${\cal B}_{ij}$ is the
``magnetic type'' tidal field of the external universe and ${\cal S}_{ij}$ is
the body's current quadrupole moment \cite{Zhang31}.  In this paper we confine
attention to the leading-order term. 
} 
\begin{equation}
{dW\over dt} = - {1\over 2} {\cal E}_{ij} {d{\cal I}_{ij}\over dt}\;;
\label{eq:work}
\end{equation}
and he argued that this is true in general relativity theory as 
well as in Newtonian theory.
In Sec.\ II we shall give a Newtonian demonstration of this claim.  In Sec.\
III we shall give a relativistic demonstration---showing, more specifically,
that although the quadrupole/tidal interaction energy is gauge dependent,
the
work done is gauge invariant and has the value (\ref{eq:work}).

\section{Newtonian Analysis}

In this section, we consider a Newtonian body, with weak internal gravity
$| \Phi_{\rm o} | \ll 1$ (where $\Phi_{\rm o}$ is the body's gravitational
potential), subjected to an external Newtonian 
tidal field.  We assume that the external field is nearly homogeneous 
in the vicinity of the body, ${\cal L} \gg R$ (cf. Fig.\ 1; in 
the Newtonian case, the inner boundary of the vacuum ``local asymptotic 
rest frame"\footnote{We shall discuss the concept of ``local asymptotic
rest frame" near the beginning of Sec.\ III.}---indicated by the inner 
dashed circle---would be at the surface of 
the body).

In our analysis, we will consider a variety of contributions to the
total energy inside a sphere which encompasses the 
body and whose boundary lies within the local asymptotic rest frame---i.e., 
the region where the external field is nearly homogeneous
(again, cf. Fig. 1).  We denote by ${\cal V}$ the interior of this 
sphere and by $\partial {\cal V}$ its boundary.  Of greatest interest 
will be the interaction energy
(between the body and the external tidal field) and the
work done by the tidal field on the body.  Both of these quantities 
are the result of slow changes of the tidal field ${\cal E}_{jk}$ and 
the body's quadrupole moment ${\cal I}_{jk}$.

As a foundation for our analysis, consider a fully isolated system that 
includes the body of interest and other ``companion" bodies, which produce 
the tidal field ${\cal E}_{jk}$ that the body experiences.  For simplicity, 
assume that all the bodies are made of perfect fluid (a restriction that 
can easily be abandoned).  Then, for the full system, the Newtonian
gravitational energy density and energy flux can be written as

\begin{eqnarray}
\Theta^{00}_1 &=& \rho \left( \frac{1}{2} v^2 + \Pi + \Phi \right) 
                  + \frac{1}{8\pi} \Phi_{,j} \Phi_{,j} \;,
\label{eq:theta00}
\\
\Theta^{0j}_1 &=& \rho v^j \left( \frac{1}{2} v^2 + \Pi + \frac{p}{\rho} 
                  + \Phi \right)
                  - \frac{1}{4\pi} \Phi_{,t} \Phi_{,j} \;,
\label{eq:theta0j}
\end{eqnarray}
where $\Phi, \rho, p, v,$ and $\Pi$ are the Newtonian gravitational 
potential, mass density, pressure, velocity, and specific internal energy
\cite{TidalHeating}.  

Using conservation of rest mass 
$\rho_{,t} + (\rho v^j)_{,j} = 0$, the first law of thermodynamics 
$\rho {d\Pi}/{dt} + pv^j_{\ ,j} =0$, the Euler equation for fluids 
$\rho {dv^j}/{dt} + \rho \Phi_{,j} +p_{,j} =0$, Newton's field 
equation $\Phi_{,jj}=4\pi \rho$, and the definition of the comoving time
derivative ${d}/{dt} = {\partial}/{\partial t} + 
v^j {\partial}/{\partial x^j}$, it can be shown that Eqs.\ 
(\ref{eq:theta00}) and (\ref{eq:theta0j}) 
satisfy conservation of energy

\begin{equation}
{\Theta^{00}}_{,t} + {\Theta^{0j}}_{,j} = 0.
\label{eq:divtheta}
\end{equation}

Equations (\ref{eq:theta00}) and (\ref{eq:theta0j}) for the energy density 
and flux, however, are not unique.  Equally valid 
are the following expressions:
\begin{eqnarray}
\Theta^{00}_2 &=&  \rho \left( \frac{1}{2} v^2 + \Pi \right) 
                   - \frac{1}{8\pi} \Phi_{,j} \Phi_{,j} \;, 
\label{eq:theta002}
\\
\Theta^{0j}_2 &=&  \rho v^j \left( \frac{1}{2} v^2 + \Pi + \frac{p}{\rho} 
                  + \Phi \right)
                  + \frac{1}{4\pi} \Phi_{,tj} \Phi \;,
\label{eq:theta0j2}
\end{eqnarray}
which also satisfy energy conservation (\ref{eq:divtheta}) but localize the 
gravitational energy in a different manner from $\Theta^{0 \mu}_1$.  Energy
conservation (\ref{eq:divtheta}) will also be satisfied by any linear 
combination of 
$\Theta^{0 \mu}_1$ and $\Theta^{0\mu}_2$.  Imposing the additional 
condition that, for any acceptable $\Theta$, the system's total energy
\begin{equation}
E = \int \Theta^{00} d^3x
\label{eq:NtotalEnergy}
\end{equation}
must be independent of the choice of $\Theta^{00}$ forces the 
coefficients to sum to 1.  Hence, a perfectly valid form for
$\Theta^{0 \mu}$ is 
\begin{eqnarray}
\Theta^{0\mu} &=&  \alpha \Theta^{0\mu}_1 + (1-\alpha ) \Theta^{0\mu}_2 \;,
\label{eq:NGenGauge}
\end{eqnarray}
where $\alpha$ is an arbitrary constant.

Notice that the choice of $\alpha$ gives a specific energy localization.
For example, $\alpha=0$ puts the gravitational energy entirely in the field 
[$-(\nabla \Phi)^2 /(8\pi)$], so it is non-zero outside the matter.  
This is analogous
to the localization used in electrostatics ($1/8\pi$ times the square of 
the electric field).  Choosing 
$\alpha=\frac{1}{2}$, by contrast, puts the gravitational energy 
entirely in the matter 
($\frac{1}{2} \rho \Phi$), so it vanishes outside the body.  When 
$\frac{1}{2} \rho \Phi$ is 
integrated over the entire system (body plus its companions), the 
result is a widely used way of computing gravitational energy (e.g., 
Sec.\ 17.1 of Ref. \cite{CoxGiuli}). 

The energy in the region ${\cal V}$ that contains and surrounds the 
body but excludes the companion,
\begin{equation}
E_{\cal V} = \int_{\cal V} \Theta^{00} d^3x \;,
\label{eq:NEnergy}
\end{equation}
will
depend on $\alpha$; i.e., it will depend on where the 
energy is localized.  By contrast, the total energy (\ref{eq:NtotalEnergy}) 
for the fully isolated system (body plus its companions) 
will be independent of $\alpha$. 

Another way to express this ambiguity of the localization of the 
gravitational energy is given by Thorne 
(Appendix of Ref. \cite{TidalHeating}): it is possible to add the divergence of 
$\eta_j = \beta \Phi \Phi_{,j}$ (where $\beta$ is an arbitrary constant) 
to $\Theta^{00}_1$ and the time derivative of $-\eta_j$ to $\Theta^{0j}_1$
without affecting energy conservation (\ref{eq:divtheta}) or the 
physics of the system.  Indeed, this method is completely equivalent 
to the one presented above.  The constants are related by 
$\beta = ( \alpha -1) / 4\pi$.

Throughout the region $\cal V$, the Newtonian 
gravitational potential can be broken into two parts:  the body's self field
$\Phi_{\rm o}$ and the tidal field $\Phi_{\rm e}$ produced by the external
universe (the companion bodies), so that
\begin{equation}
\Phi = \Phi_{\rm o} + \Phi_{\rm e} \;.
\label{eq:PhiSum}
\end{equation}
The external field is quadrupolar and source-free in the region $\cal V$ so that
\begin{equation}
\Phi_{\rm e} = {1\over2} {\cal E}_{ij} x^i x^j \;,  \quad
\Phi_{{\rm e},jj} = 0 \;, 
\label{eq:Phie}
\end{equation}
and, furthermore, the tidal field ${\cal E}_{ij}$ evolves slowly with time.  The 
body's (external) self field is monopolar and quadrupolar and has the body's 
mass distribution as a source:
\begin{equation}
\Phi_{\rm o} = - {M\over r} -{3\over2} {{\cal I}_{ij}n^i n^j \over r^3} \;, \quad 
\Phi_{{\rm o},jj} = 4\pi\rho\;.
\label{eq:Phio}
\end{equation}
The quadrupole moment ${\cal I}_{ij}$, like that of the external field, evolves
slowly with time, but the body's mass $M$ is constant.  Recall that 
$r \equiv \sqrt{\delta_{ij} x^i x^j}$ is radial distance from the body's
center of mass and 
$n^j \equiv x^j/r$ is the radial vector.

A useful expression for the total energy $E_{\cal V}$ in the 
spherical region ${\cal V}$ can be derived by inserting 
Eqs.\ (\ref{eq:NGenGauge}), (\ref{eq:theta00}), (\ref{eq:theta002}), 
and (\ref{eq:PhiSum}) into 
Eq.\ (\ref{eq:NEnergy}).  The resulting expression can be broken into 
a sum of three parts---the body's self energy $E_{\rm o}$ (which depends
only on $\Phi_{\rm o}$ and $\rho$), 
the external field energy $E_{\rm e}$ (which depends only on $\Phi_{\rm e}$), 
and the interaction energy 
$E_{\rm int}$ (which involves products of $\Phi_{\rm e}$ with $\rho$ or
$\Phi_{\rm o}$):

\begin{equation}
E_{\cal V} = E_{\rm o} + E_{\rm e} + E_{\rm int} \;, 
\label{eq:Esum}
\end{equation}
where
\begin{eqnarray}
E_{\rm o} = \int_{\cal V} \biggl[ && \rho \left( \frac{1}{2}v^2 + \Pi \right) 
          + \alpha \rho \Phi_{\rm o} \nonumber \\
          && + \frac{(2\alpha-1)}{8\pi} \Phi_{{\rm o},j} \Phi_{{\rm o},j} 
          \biggr] d^3x \;, \label{eq:NEo} \\
E_{\rm e} =  \int_{\cal V} \biggl[ &&
           \frac{(2\alpha-1)}{8\pi} \Phi_{{\rm e},j} \Phi_{{\rm e},j} 
           \biggr] d^3x \;, \label{eq:NEe} \\
E_{\rm int} =  \int_{\cal V}  \biggl[ && \alpha \rho \Phi_{\rm e} 
          + \frac{(2\alpha-1)}{4\pi} \Phi_{{\rm o},j} \Phi_{{\rm e},j} 
          \biggr] d^3x \;. \label{eq:NEintf}
\end{eqnarray}
Inserting Eqs.\ (\ref{eq:Phie}) and (\ref{eq:Phio}) into Eq.\ (\ref{eq:NEintf}) 
and integrating gives the interaction energy inside $\cal V$ as

\begin{equation}
E_{\rm int} = \frac{(2+\alpha)}{10} {\cal E}_{ij} {\cal I}_{ij}\;,
\label{eq:neint}
\end{equation}
which depends on $\alpha$.  In other words, it depends on our 
arbitrary choice of how to localize
gravitational energy.  This is the ambiguity of the interaction energy
discussed in Sec.\ I.

The rate of change of the total energy inside $\cal V$ can be 
expressed in the form 

\begin{equation}
\frac{dE_{\cal V}}{dt} = - \int_{\partial {\cal V}} \Theta^{0j} n^j r^2 d\Omega
\end{equation}
by taking the time derivative of Eq.\ (\ref{eq:NEnergy}), inserting 
Eq.\ (\ref{eq:divtheta}), and applying the divergence theorem.  
This expression, like that for the energy, can be broken into a sum
by combining Eqs.\ (\ref{eq:NGenGauge}), (\ref{eq:theta0j}), (\ref{eq:theta0j2}),
 (\ref{eq:PhiSum}), (\ref{eq:Phie}), (\ref{eq:Phio}), and (\ref{eq:neint}):  

\begin{eqnarray}
\frac{dE_{\cal V}}{dt} && = \frac{dE_{\rm e}}{dt} 
 		     + \frac{dE_{\rm int}}{dt}
                     - \frac{1}{2} {\cal E}_{ij} \frac{d{\cal I}_{ij}}{dt} 
                     \nonumber \\ 
                 + && \int_{\partial {\cal V}} \frac{1}{4\pi} \left[
                     \alpha \Phi_{{\rm o},t} \Phi_{{\rm o},j} 
                     + \left( \alpha - 1 \right) \Phi_{{\rm o},tj} \Phi_{\rm o} 
                     \right] n^j r^2 d\Omega \;.
\label{eq:dEdtx}
\end{eqnarray}
The first term is the rate of change of the external field energy 
(\ref{eq:NEe}) inside
${\cal V}$, resulting from the evolution of the tidal field.  The second
term is the rate of change of the interaction energy (\ref{eq:neint}).
The third and fourth terms together, by comparison with Eq.\ 
(\ref{eq:Esum}), must 
be equal to $dE_{\rm o}/dt$, the rate of change of the body's self energy.
The fourth term is the contribution from the
body's own field moving across the boundary $\partial {\cal V}$.  Therefore,
the third term gives the 
change in the body's energy coming from the interaction with 
the tidal field; in other words, it is the rate of work done 
on the body by the tidal field.  Furthermore, this term is 
independent of $\alpha$ and is, consequently, independent of how the
Newtonian energy is localized,
as claimed in Sec.\ I, Eq.\ (\ref{eq:work}).

\section{Relativistic Analysis}

In this section, we will exhibit a relativistic version of the calculation in
Sec.\ II, again showing that the rate of work done on the 
body by the tidal field is 
gauge invariant and that it has the same value in a 
 general relativistic perturbative treatment 
as in the Newtonian one:  
$-(1/2) {\cal E}_{ij} d{\cal I}_{ij}/dt$.  
The formalism to be used is the Landau-Lifshitz 
energy-momentum pseudotensor and multipole expansions as developed by
Thorne and Hartle \cite{ThorneHartle} and Zhang \cite{Zhang34,Zhang31}, 
together with the slow-motion approximation, so time derivatives are 
small compared to spatial gradients.  

We will work in the body's vacuum local asymptotic rest frame, which is 
defined as a region outside the body and far enough from it that its
gravitational field can be regarded as a weak perturbation of flat 
spacetime, yet near enough that the tidal field of the external universe 
can be regarded as nearly homogeneous.  This
region is a spherical shell around the body; its inner boundary is near the 
body's surface but far enough away for gravity to be weak, and its outer 
boundary is at a distance where the external 
tidal field begins to depart from homogeneity (see Fig. 1).  Somewhat more 
precisely, the local asymptotic rest frame is the region throughout which
$r/{\cal L} \ll 1$, $r/{\cal R} \ll 1$, and $M/r \ll 1$, where $r$ is 
the radial distance from the 
body, $M$ is the mass of the body, and $\cal R$ and ${\cal L}$
are the radius of curvature and the scale of homogeneity of the external 
gravitational field.  If this region exists (as in the case of a black hole
binary far from merger, for example) and the slow-motion limit applies, then
the following analysis is valid.

As in the Newtonian case, we will consider contributions to the total energy
inside a sphere $\cal V$ which encompasses the body and whose boundary 
$\partial \cal V$ lies within the local asymptotic rest frame (see Fig. 1).

\subsection{deDonder Gauge}

We shall begin by computing the work done on the body by the external 
tidal field, using deDonder gauge (this section).  Then in the next 
section, we shall show that the work is gauge invariant.

DeDonder gauge is defined
by the condition that ${\bar h}^{\alpha \beta}_{\ \ ,\beta} =0$, 
where ${\bar h}^{\alpha \beta}$ is defined in terms of the metric density 
as follows:
\begin{equation}
{\frak g}^{\alpha \beta} \equiv \sqrt{-g} \ g^{\alpha \beta} \equiv 
        \eta^{\alpha \beta} - \bar{h}^{\alpha \beta} \;.
\label{eq:metric}
\end{equation}
At linear order in the strength of gravity, $\bar{h}^{\alpha \beta}$ is the 
trace-reversed metric perturbation.  According to Zhang \cite{Zhang31},
the components of $\bar{h}^{\alpha \beta}$ in the body's local asymptotic
rest frame
are, at leading (linear) order in the strength of gravity and at leading 
non-zero order in our slow-motion expansion,

\begin{eqnarray}
\bar{h}^{00} && \equiv -4 \Phi = 4 {M \over r} 
                               + 6 {{{\cal I}_{ij} x^i x^j}\over r^5} 
                               - 2 {\cal E}_{ij} x^i x^j , \label{eq:h00d}\\
\bar{h}^{0j} && \equiv -A_j = 6 {{\dot {\cal I}}_{ja} x^a \over r^3}
                                 + {10\over 21} {\dot{\cal E}}_{ab} x^a x^b x^j 
                                 - {4 \over 21} {\dot {\cal E}}_{ja} x^a r^2 ,
                                 \label{eq:h0jd} \\
\bar{h}^{ij} && = {\cal O} \left( 
      \ddot{\cal E}_{ij} r^4 \  \& \  \frac{\ddot{\cal I}_{ij}}{r} \right) ,
      \label{eq:hijd}                               
\end{eqnarray}
where the dots indicate time derivatives (i.e., ${\dot {\cal I}}_{ij} \equiv
 d{\cal I}_{ij}/dt$) and the symbol ``\&" means ``plus terms of the form 
 and magnitude."  
Note that the higher-order ($\ell$-order) multipoles have been omitted, since 
their contributions are smaller by 
$\sim (r/{\cal L})^{\ell-2} \leq r/{\cal L} \ll 1$ and 
$\sim (M/r)^{\ell-2} \leq M/r \ll 1$ than the quadrupolar ($\ell =2$)
terms that we have kept.  The second 
time derivatives will also be neglected since they are unimportant 
in the slow-motion approximation;
this effectively eliminates $\bar{h}^{ij}$ in this gauge.  Also, note that
the $\Phi$ in Eq.\ (\ref{eq:h00d}) has the same form as the Newtonian 
gravitational potential of
Sec.\ II, and the $A_j$ of Eq.\ (\ref{eq:h0jd}) is a gravitational vector
potential, which does not appear in Newtonian theory.

In general, it would be possible to have a term 
$\propto {\cal I}_{jk} {\cal E}_{jk} /r$ 
in Eq.\ (\ref{eq:h00d}), as well as the $\propto M/r$ term.  We have chosen 
to define the 
constant for the monopolar term to be $4M$, thereby eliminating any term
$\propto {\cal I}_{jk} {\cal E}_{jk} /r$; this is arbitrary but convenient,
as will be seen shortly.

The total mass-energy $\cal M_V$ inside the sphere $\cal V$ 
(total stellar mass including the quadrupolar deformation 
energy and energy of interaction between the deformation and tidal field)
is defined by Thorne and Hartle (Eq.\ (2.2a) of Ref.
\cite{ThorneHartle}) in terms of the Landau-Lifshitz superpotential as

\begin{equation}
{\cal M_V} = \frac{1}{16\pi} \int_{\partial {\cal V}} 
H^{0\alpha 0j}_{\ \ \ \ ,\alpha} n_j r^2 d\Omega \;,
\label{eq:massf}
\end{equation}
where
\begin{equation}
H^{\mu \alpha \nu \beta} = {\frak g}^{\mu \nu} {\frak g}^{\alpha \beta} - 
{\frak g}^{\alpha \nu} {\frak g}^{\mu \beta} \;;
\label{eq:H}
\end{equation}
cf. Eqs.\ (20.6), (20.3), and (20.20) of MTW \cite{MTW} and 
Eqs.\ (100.14) and (100.2) of Landau and Lifshitz \cite{LanLif}.
Using Eqs.\ (\ref{eq:H}), (\ref{eq:metric}), and 
(\ref{eq:h00d})--(\ref{eq:hijd}) in Eq.\ (\ref{eq:massf}) and
carrying out the integration gives
\begin{eqnarray}
{\cal M_V} = M &&+ {\cal O} \left( {{\cal E}} {\ddot{\cal E}} r^7 \ \& \  
				{\dot{\cal E}} {\dot{\cal E}} r^7 \right) 
				\nonumber \\
	     &&+ {\cal O} \left( {\dot{\cal I}} {\dot{\cal E}} r^2 \ \& \  
				{\ddot{\cal I}} {{\cal E}} r^2 \  \& \ 
				{{\cal I}} {\ddot{\cal E}} r^2 \right)
				\nonumber \\
	     &&+ {\cal O} \left( \frac{{{\cal I}} {\ddot{\cal I}}}{r^3} \ \& \  
				\frac{{\dot{\cal I}} {\dot{\cal I}}}{r^3} 
				\right)\;.
\label{eq:dMvres}
\end{eqnarray}
Hence, ${\cal M_V} = M$ at leading order in our slow-motion approximation
when we neglect the double and higher-order 
time derivatives.  This is the reason the constant of the monopolar term in 
Eq.\ (\ref{eq:h00d}) was chosen to be $4M$.

To calculate the rate of work done by the tidal field on the body when 
${\cal E}_{jk}$ and ${\cal I}_{jk}$ change slowly, we 
consider the change of the mass-energy
${\cal M_V}$,
\begin{eqnarray}
-{d{\cal M_V} \over dt} &=& 
	\int_{\partial {\cal V}} (-g) t^{0j} n_j r^2 d\Omega \;,
\label{eq:dMdt}
\end{eqnarray}
where $t^{\mu \nu}$ is the Landau-Lifshitz pseudotensor.  That this 
expression is indeed the time derivative of Eq.\ (\ref{eq:massf}) is a
result of Gauss's theorem (see discussion after Eqs.\ (2.3) of Ref.
\cite{ThorneHartle}).

In the body's local asymptotic rest frame, the Landau-Lifshitz energy-momentum 
pseudotensor (Eq.\ (20.22) in MTW \cite{MTW}) is, at the orders of accuracy we
are considering, 

\begin{eqnarray}
(-g)t^{00} = && {1 \over 16\pi} \left( -{7 \over 8} \right) g^{ij} 
                  \bar{h}^{00}_{\ \ ,i} \bar{h}^{00}_{\ \ ,j}  \nonumber \\ 
           = && -{7 \over 8\pi} \delta^{ij} \Phi_{,i} \Phi_{,j} \;, 
\label{eq:t00} \\
(-g)t^{0j} = && {1 \over 16\pi} \Biggl( 
                   {3 \over 4} \bar{h}^{00}_{\ \ ,0} \bar{h}^{00}_{\ \ ,j} 
                   + \bar{h}^{00}_{\ \ ,m} \bar{h}^{0m}_{\ \ ,j} 
                   - \bar{h}^{00}_{\ \ ,m} \bar{h}^{0j}_{\ \ ,m}   
                \Biggr) \nonumber \\ 
           = && {3 \over 4\pi} \Phi_{,0} \Phi_{,j} +
                 {1 \over 4 \pi} \left( A_{k,j} - A_{j,k}\right) 
                 \Phi_{,k} \;.
\label{eq:t0j}
\end{eqnarray}
Here the deDonder gauge condition ${\bar h}^{\alpha \beta}_{\ \ ,\beta} = 0$
has been used to eliminate many terms from the general expression in MTW, 
but most of the
simplification has come from keeping only terms of leading-order in the
slow-motion approximation
[zeroth and first time derivatives, respectively, in Eqs.\ (\ref{eq:t00}) 
and (\ref{eq:t0j})].  This restriction
has given us only products of 
${\bar h}^{\mu \nu}_{\ \ ,\alpha}$ which will produce terms containing the 
products 
$M^2,\ M {\cal E},\ M {\cal I},\ {\cal I}{\cal E},\ {\cal I}{\cal I},\ 
{\cal E}{\cal E}$ for 
$(-g)t^{00}$ and $M {\dot {\cal I}},\ M {\dot {\cal E}},\ 
{\cal I}{\dot{\cal E}},\ {\cal E}{\dot{\cal I}},\ 
{\cal I}{\dot {\cal I}},\ {\cal E}{\dot {\cal E}}$ for $(-g)t^{0j}$.
  This may be illustrated
by expressing $(-g)t^{0\alpha}$ explicitly in terms of the quadrupole moments by 
substituting Eqs.\
(\ref{eq:h00d}) and (\ref{eq:h0jd}) into Eqs.\ (\ref{eq:t00}) and 
(\ref{eq:t0j}) to get

 \begin{eqnarray}
 (-g)t^{00} && = {1 \over 16\pi} \Biggl( 
 			- 14 {M^2 \over r^4} 
 			- 210 {{\cal I}_{ab} {\cal E}_{cd} x^a x^b x^c x^d 
 				\over r^7} \nonumber \\
 		        + && 84 {{\cal I}_{ab} {\cal E}_{bc} x^a x^c \over r^5}
 			- 28 {M {\cal E}_{ab} x^a x^b \over r^3} 
 			- 14 {\cal E}_{ab} {\cal E}_{bc} {x^a x^c} \nonumber \\
 		        - && 126 {M {\cal I}_{ab} x^a x^b \over r^8}
 		        - 126 {{\cal I}_{ab} {\cal I}_{bc} x^a x^c \over r^{10}}
 		        	\nonumber \\
 		        - && {315 \over 2} {{\cal I}_{ab} {\cal I}_{cd} 
 				x^a x^b x^c x^d \over r^{12}} \Biggr) \;,
\end{eqnarray}
and
\begin{eqnarray} 
(-g)t^{0j} = {1 \over 16\pi} \Biggl( && 
                - 18 {{\dot {\cal I}}_{ab} {\cal E}_{cj} x^a x^b x^c \over r^5} 
                + 24 {{\dot {\cal I}}_{ab} {\cal E}_{bc} x^a x^c x^j \over r^5} 
                	\nonumber \\
            &&  - 24 {{\dot {\cal I}}_{aj} {\cal E}_{bc} x^a x^b x^c \over r^5} 
                - 18 {{\cal I}_{aj} {\dot {\cal E}}_{bc} x^a x^b x^c \over r^5} 
                	\nonumber \\
            &&  - 24 {{\cal I}_{ab} {\dot {\cal E}}_{cj} x^a x^b x^c \over r^5}
                - 16 {{\cal I}_{ab} {\dot {\cal E}}_{bc} x^a x^c x^j \over r^5} 
                	\nonumber \\
            &&  + 85 {{\cal I}_{ab} {\dot {\cal E}}_{cd} 
                	x^a x^b x^c x^d x^j \over r^7} \Biggr) \;.
\label{eq:t0jquad}
\end{eqnarray}
Note that we have kept only the $\cal EI$ cross terms in the expression for 
$(-g) t^{0j}$, as only they will 
contribute to our calculation of the interaction energy and work.  

We find the rate of change of mass-energy inside our  
sphere $\cal V$ by inserting Eq.\ (\ref{eq:t0jquad}) 
into Eq.\ (\ref{eq:dMdt}) and integrating.  The result (only considering the
cross terms) is

\begin{eqnarray}
-{d{\cal M_V} \over dt}   
           &=& {d \over dt}\left( \frac{1}{10} {\cal E}_{ij} 
           	{\cal I}_{ij} \right) +
             \frac{1}{2} {\cal E}_{ij} \frac{d}{dt} {\cal I}_{ij} \;.
\label{eq:Dresult}
\end{eqnarray}
Since the interaction energy can depend
only on the instantaneous deformation and tidal field, its rate of change
must be a perfect differential, whereas the rate of work need not be.  
Also, if the tidal field changes but the body does not, there is no work 
done.  From these two facts, we can conclude that the first term of Eq.\
(\ref{eq:Dresult}) is the rate of change of the interaction energy 
between the tidal field and the body and that the second term 
represents the rate of work done by the 
external field on the body (the ``tidal heating'').  

Notice that this value for the rate of work matches that discussed 
in Sec.\ I, Eq.\ (\ref{eq:work}), and found via the Newtonian analysis 
in Sec.\ II.  Note the comparison with our Newtonian expressions 
(\ref{eq:dEdtx}) and (\ref{eq:neint}).  The first and
fourth terms of Eq.\ (\ref{eq:dEdtx}) are missing here because we 
included in our $(-g)t^{0j}$ only the (body field)$\times$(external field) 
crossterms.  If we had also included (external)$\times$(external) and 
(body)$\times$(body) terms, Eq.\ (\ref{eq:Dresult}) would have entailed 
expressions like the 
first and fourth terms of Eq.\ (\ref{eq:dEdtx}).  Note also from the 
interaction energy terms in Eqs.\ (\ref{eq:dEdtx}), (\ref{eq:neint}),
and (\ref{eq:Dresult}) 
that the Landau-Lifshitz way of localizing gravitational energy corresponds
to the Newtonian choice $\alpha = -3$, a correspondence that has previously
been derived by Chandrasekhar \cite{Chandra}.
  
 \subsection{General Gauge}
 
 In Sec.\ III.\ A., we considered the special case of deDonder gauge, which was
 particularly simple due to the gauge condition and to the ${\bar h}^{ij}$ 
 terms being effectively zero.  Now we will examine a general gauge, which can 
 be achieved by a gauge transformation of the form
 
\begin{eqnarray}
{\bar h}_{\mu\nu} \rightarrow && {\bar h}_{\mu\nu} + \delta {\bar h}_{\mu\nu} \;;
	\nonumber \\
\delta {\bar h}_{\mu\nu} = 
\xi_{\mu,\nu} && + \xi_{\nu,\mu} - \eta_{\mu \nu} \xi^\alpha_{\ ,\alpha} \;,
\label{eq:gaugechange}
\end{eqnarray}
where $\xi^\alpha$ is a function to be chosen shortly.
Using ${\bar h}_{\mu \nu} \equiv h_{\mu \nu} - \frac{1}{2} \eta_{\mu \nu} 
h^\sigma_{\ \sigma}$, where $h_{\mu \nu}$ is the perturbation of the metric
away from the Minkowski metric in the local asymptotic rest frame, 
this can also be expressed as
\begin{eqnarray}
&& {h}_{\mu\nu} \rightarrow {h}_{\mu\nu} + \delta {h}_{\mu\nu} \;; \nonumber \\
&& \delta {h}_{\mu\nu} = \xi_{\mu,\nu} + \xi_{\nu,\mu} \;.
\end{eqnarray}
Note that in the deDonder gauge, 
\begin{eqnarray}
h_{00} && = 2 {M \over r} + 3 {{{\cal I}_{ij} x^i x^j}\over r^5} 
            - {\cal E}_{ij} x^i x^j , \label{eq:h00g}\\
h_{0j} && = - 6 {{\dot {\cal I}}_{ja} x^a \over r^3}
            - {10\over 21} {\dot{\cal E}}_{ab} x^a x^b x^j 
            + {4 \over 21} {\dot {\cal E}}_{ja} x^a r^2 ,
            \label{eq:h0jg} \\
h_{ij} && = \delta_{ij} \left( 2 {M \over r} 
    	        + 3 {{{\cal I}_{k \ell} x^k x^\ell}\over r^5} 
     	        - {\cal E}_{k \ell} x^k x^\ell
       	     \right) . \label{eq:hijg}                               
\end{eqnarray}

Since we are interested only in gauge changes of the same order as we have 
been using so far (leading-order in the slow-motion approximation), 
we include only 
terms in $\xi_\alpha$ that will produce $\delta h_{\mu \nu}$ of the same 
forms as Eqs.\ (\ref{eq:h00g})--(\ref{eq:hijg}), but with different 
numerical coefficients.  For example, consider
$\delta h_{00} =2 \xi_{0,0} \propto M/r$; that gives $\xi_0 \propto Mt/r$, since 
$M$ is a constant.  This, in turn, implies
$\delta h_{0j} = \xi_{0,j} \propto Mt x^j/r^3$, but this is not of the
same form as the terms in Eq.\ (\ref{eq:h0jg}); rather, it corresponds to a 
gauge that rapidly becomes ill-behaved as time passes.  Similar arguments apply
to $\delta h_{00} \propto {\cal I}_{jk} x^j x^k/r^5$ or 
$\delta h_{00} \propto {\cal E}_{jk} x^j x^k$; their coefficients cannot be 
altered  by a gauge change because such a change would alter the mathematical
form of $h_{0j}$ and would make its magnitude unacceptably large in the 
slow-motion limit.  As a result, we must set $\xi_0 = 0$.
If we now consider $\delta h_{0j} = \xi_{j,0}$, possible terms are of the
form $ \propto {\dot{\cal I}}_{ja} x^a /r^3$, 
$\propto {\dot {\cal E}}_{ab} x^a x^b x^j$, or
$\propto {\dot {\cal E}}_{ja} x^a r^2$; cf. Eq.\ (\ref{eq:h0jg}).  Each of 
these gives $\delta h_{00} = 0$ and $\delta h_{ij}$ of the same form 
as Eq.\ (\ref{eq:hijg}); hence, terms of these forms are allowed.  
Consequently, the most general gauge change
that preserves the mathematical forms of $h_{\mu \nu}$ but alters their 
numerical coefficients is
   
 \begin{eqnarray}
 \xi_0 &=& 0 \;, \nonumber \\
 \xi_j &=& \alpha {{\cal I}_{ja} x^a \over r^3} + \beta {\cal E}_{ja} x^a r^2 +
           \gamma {\cal E}_{ab} x^a x^b x^j \;,
 \label{eq:xi}
 \end{eqnarray}
 where $\alpha, \beta,$ and $\gamma$ are arbitrary constants.
 
 Using the trace-reversed gauge change (\ref{eq:gaugechange}) with 
 Eq.\ (\ref{eq:xi}), the new $\bar{h}^{\alpha \beta}$ become
 
\begin{eqnarray}
\bar{h}^{00} = && 4{M\over r} +(5 \gamma +2\beta - 2) {\cal E}_{ij} x^i x^j
			\nonumber \\
               && + (6-3\alpha) {\cal I}_{ij} {{x^i x^j}\over r^5}  \;,
\label{eq:h00} \\
\bar{h}^{0j} = && (2-\alpha) {\dot {\cal I}}_{ja} {x^a \over r^3} + 
                ({10\over 21}-\gamma) {\dot {\cal E}}_{ab} x^a x^b x^j 
                	\nonumber \\
               && - ({4 \over 21} +\beta) {\dot {\cal E}}_{ja} x^a r^2 \;, \\                 
\bar{h}^{jk} = && 2 \alpha {\cal I}_{jk} {1 \over r^3} 
		+3 \alpha \delta^{jk} {\cal I}_{ab} \frac{x^a x^b}{r^5}
                -3 \alpha {\cal I}_{ja} \frac{x^a x^k}{r^5} \nonumber \\
             && -3 \alpha {\cal I}_{ka} \frac{x^a x^j}{r^5}
                +2 \beta {\cal E}_{jk} r^2
                -2 ( \beta + \gamma ) \delta^{jk} {\cal E}_{ab} x^a x^b 
                	\nonumber \\
             && +2 ( \beta + \gamma ) {\cal E}_{ja} x^a x^k 
                +2 ( \beta + \gamma ) {\cal E}_{ka} x^a x^j \;.
\label{eq:hjk}
\end{eqnarray}

In this general gauge, if we calculate the total mass-energy ${\cal M_V}$ 
inside the sphere $\cal V$ using Eqs.\ (\ref{eq:H}), (\ref{eq:metric}), and
(\ref{eq:h00})--(\ref{eq:hjk}) in Eq.\ (\ref{eq:massf}), 
we find

\begin{eqnarray}
{\cal M_V} = && M + \left( 2\gamma^2 + \frac{29}{5} \beta \gamma 
			- \frac{4}{5} \gamma + 2 \beta^2 -2\beta \right)
			 {\cal E}_{ij} {\cal E}_{ij} r^5 \nonumber \\
               && + \left( \frac{1}{3} \alpha + 2\beta + \frac{7}{5} \gamma 
             		- \frac{4}{3} \alpha \beta 
             		- \frac{23}{15} \alpha \gamma \right) 
             		{\cal I}_{ij} {\cal E}_{ij} \;.
\label{eq:Mvresult}
\end{eqnarray}
The ${\cal I}_{ij}{\cal I}_{ij}/r^5$ term is zero.   
It is comforting to note that all the new terms vanish for 
$\alpha = \beta = \gamma = 0$, giving the deDonder result ${\cal M_V} = M$.
The ${\cal E E} r^5$ and ${\cal I E}$ terms in Eq.\
(\ref{eq:Mvresult}) are
large compared to the double time-derivative terms that formed the largest 
corrections to the mass-energy in deDonder gauge (\ref{eq:dMvres}); however,
they are still small compared to the mass $M$ that appears in the expansion 
(\ref{eq:h00d}) of $\bar{h}^{00}$.  
Also note that, near the body of interest, the ${\cal E E} r^5$ term
will be small compared to the $\cal I E$ term, due to its radial dependence.
So, once again,
we have ${\cal M_V} \simeq M$ as a first approximation, although it is 
necessary to keep the extra terms in Eq.\ (\ref{eq:Mvresult}) to maintain the
same level of accuracy as we have been using.  Consequently, 
$\cal M_V$ is gauge-dependent to the order that interests us, and it has
 the ``${\cal I}_{ij} {\cal E}_{ij}$" ambiguity discussed by Thorne 
 and Hartle \cite{ThorneHartle}
and mentioned in Sec.\ I.

Keeping only the leading-order terms in the slow-motion approximation, 
as described in Sec.\ III.\ A., 
the Landau-Lifshitz pseudotensor in the new gauge is 

\begin{eqnarray}
(-g)t^{00} = {1 \over 16\pi} \Biggl( && 
                   - {7 \over 8} g^{ij} \bar{h}^{00}_{\ \ ,i} 
                   	\bar{h}^{00}_{\ \ ,j} 
                   + \bar{h}^{00}_{\ \ ,i} \bar{h}^{ij}_{\ \ ,j} \nonumber \\ 
                && - \frac{1}{2} \bar{h}^{ij}_{\ \ ,k} \bar{h}^{ik}_{\ \ ,j} 
                   + \frac{1}{4} \bar{h}^{00}_{\ \ ,i} \bar{h}^{jj}_{\ \ ,i}
                   	\nonumber \\	
                && + \frac{1}{4} \bar{h}^{ij}_{\ \ ,k} \bar{h}^{ij}_{\ \ ,k}    
                   - \frac{1}{8} \bar{h}^{ii}_{\ \ ,k} \bar{h}^{jj}_{\ \ ,k} 
                   \Biggr) \;,
\label{eq:t00gen}
\end{eqnarray}

\begin{eqnarray}                
(-g)t^{0j} =  {1 \over 16\pi}  \Biggl( &&
                   {3 \over 4} \bar{h}^{00}_{\ \ ,0} \bar{h}^{00}_{\ \ ,j} 
                   + \bar{h}^{00}_{\ \ ,m} \bar{h}^{0m}_{\ \ ,j} 
                   - \bar{h}^{00}_{\ \ ,m} \bar{h}^{0j}_{\ \ ,m} \nonumber \\
            &&     + \bar{h}^{0j}_{\ \ ,k} \bar{h}^{kl}_{\ \ ,l}   
                   - \bar{h}^{00}_{\ \ ,0} \bar{h}^{ij}_{\ \ ,j}
                   + \bar{h}^{ij}_{\ \ ,k} \bar{h}^{ik}_{\ \ ,0} \nonumber \\
            &&     - \bar{h}^{0i}_{\ \ ,k} \bar{h}^{ik}_{\ \ ,j} 
                   + \bar{h}^{0i}_{\ \ ,k} \bar{h}^{ij}_{\ \ ,k}    
                   - \bar{h}^{0i}_{\ \ ,i} \bar{h}^{jk}_{\ \ ,k} \nonumber \\
            &&     - \frac{1}{4} \bar{h}^{kk}_{\ \ ,0} \bar{h}^{00}_{\ \ ,j}
                   - \frac{1}{4} \bar{h}^{00}_{\ \ ,0} \bar{h}^{kk}_{\ \ ,j}  
                   	 \nonumber \\
            &&     - \frac{1}{2} \bar{h}^{ik}_{\ \ ,0} \bar{h}^{ik}_{\ \ ,j}
                   + \frac{1}{4} \bar{h}^{ii}_{\ \ ,0} \bar{h}^{kk}_{\ \ ,j} 
                   \Biggr) \;.
\label{eq:t0jgen}
\end{eqnarray}
 Note that the first term of Eq.\ (\ref{eq:t00gen}) is the same as 
 Eq.\ (\ref{eq:t00}), and the first three
 terms of Eq.\ (\ref{eq:t0jgen}) are the same as Eq.\ (\ref{eq:t0j}).  
 The additional terms all involve 
 $\bar{h}^{jk}$, which was effectively zero in the deDonder gauge because
 of our slow-motion assumption.
 
Substituting Eqs.\ (\ref{eq:t0jgen}) and (\ref{eq:h00})--(\ref{eq:hjk}) 
into Eq.\ (\ref{eq:dMdt}) and integrating 
gives the rate of change of mass-energy inside the sphere $\cal V$ as
 
 \begin{eqnarray}
-{d{\cal M_V} \over dt} 
                &=&   {d \over dt}\biggl[ 
                       \left( \frac{7 \alpha \gamma}{5} -\frac{9\gamma}{5}
                             +\frac{7 \alpha \beta}{5} -\frac{12 \beta}{5}
                             - \frac{\alpha}{5} + \frac{1}{10} \right) 
                             \nonumber \\
                  &&   \ \ \ \ \times {\cal E}_{ij} {\cal I}_{ij} \biggr] 
                       + \frac{1}{2} {\cal E}_{ij} \frac{d{\cal I}_{ij}}{dt} \;.
\label{eq:dMvdtfinal}
\end{eqnarray}

Again, we have kept only the (external field)$\times$(body field) crossterms.
As expected, this expression reduces to the deDonder result (\ref{eq:Dresult})
when $\alpha = \beta = \gamma = 0$.  Using the same argument for 
Eq.\ (\ref{eq:dMvdtfinal}) as for Eq.\ (\ref{eq:Dresult}), we can conclude that
the first term of Eq.\
(\ref{eq:dMvdtfinal}) is the rate of change of the interaction energy 
between the tidal field and the body and that the second term 
represents the rate of work done by the 
external field on the body (the ``tidal heating'').  

Notice the gauge dependence (dependence on $\alpha, \beta, \gamma$) of the
rate of change of interaction energy
\begin{eqnarray}
\frac{dE_{\rm int}}{dt} = \frac{d}{dt} \biggl[ && \left( 
			     \frac{7 \alpha \gamma}{5} -\frac{9\gamma}{5}
                             +\frac{7 \alpha \beta}{5} -\frac{12 \beta}{5}
                             - \frac{\alpha}{5} + \frac{1}{10} \right) 
                             \nonumber \\ 
            && \ \ \ \ \  \times {\cal E}_{ij} {\cal I}_{ij} \biggr] \;.
\end{eqnarray}
By contrast, the ``tidal heating" work done on the body has the same, 
gauge-invariant value as in Newtonian theory
\begin{equation}
{dW\over dt} = - {1\over 2} {\cal E}_{ij} {d{\cal I}_{ij}\over dt}\;.
\end{equation}

\section{Conclusions}

In this paper, we have shown that the rate of work done by an external tidal
field on a body is independent of how gravitational energy is localized
in Newtonian theory and that it is gauge invariant in general relativity. 
 Furthermore, this quantity---which we are calling the ``tidal 
heating"---is given unambiguously by Eq.\ (\ref{eq:work}).  That the tidal
heating should be a well-defined and precise quantity is reasonable, given 
that its physical effects have been observed in the form of volcanic 
activity of Jupiter's moon Io \cite{Peale,Ross,Ojak}.

There remains one aspect of the uniqueness of the tidal heating that we have
not explored: a conceivable (but highly unlikely) dependence of $dW/dt$ on 
the choice of energy-momentum pseudotensor in general relativity.  The 
arbitrariness of the pseudotensor is general relativity's analog of Newtonian
theory's arbitrariness of localization of gravitational energy.  The fact that
$dW/dt$ is independent of the Newtonian energy localization suggests
that it may also be independent of the general relativistic pseudotensor.  
In addition, the clear physical nature of $dW/dt$ gives further confidence 
that it must be independent of the pseudotensor.  Nevertheless, it would be
worthwhile to verify explicitly that $dW/dt$ is pseudotensor-independent.

\section*{Acknowledgments}

I thank Kip Thorne for proposing this research problem and for helpful
advice about its solution and about the prose of this paper. 
This research was supported in part by NSF Grant AST-9731698 and NASA Grant
NAG5-6840.

\begin{figure}
\caption{An example of an isolated, slow-motion body: a star or black hole 
in a binary system with $R/a \ll 1$, where $R$ is the radius of the body
and $a$ is the separation of the body and companion.  The dashed circles 
indicate the boundaries of the body's vacuum local
asymptotic rest frame, the region in which $M/r \ll 1$ and 
${r}/{\cal L} \ll 1$.  Here, $r$ is the radial coordinate, $M$ is the mass
of the body, and $\cal L$ is the scale of homogeneity of the gravitational
field.  The boundary of the sphere over which we
integrate, denoted by $\partial {\cal V}$, lies within this region.}
\end{figure}


\begin{references}

\bibitem{ThorneHartle} K.\ S.\ Thorne and J.\ B.\ Hartle, Phys.\ Rev.\ D, {\bf
31}, 1815 (1985).

\bibitem{Peale} S.\ J.\ Peale, P.\ Cassen, and R.\ T.\ Reynolds, Science 
{\bf 203}, 894 (1979).

\bibitem{MTW} C.\ W.\ Misner, K.\ S.\ Thorne and J.\ A.\ Wheeler, {\it
Gravitation}, (W.\ H.\ Freeman, San Francisco, 1973).

\bibitem{Thornermp} K.\ S.\ Thorne, Rev.\ Mod.\ Phys.\ {\bf 52}, 299 (1980).

\bibitem{Zhang34} X.-H.\ Zhang, Phys.\ Rev.\ D, {\bf 34}, 991 (1986).

\bibitem{Zhang31} X.-H.\ Zhang, Phys.\ Rev.\ D, {\bf 31}, 3130 (1985).

\bibitem{TidalHeating} K.\ S.\ Thorne, Phys.\ Rev.\ D, {\bf 58}, 124031 (1998).

\bibitem{Flanagan} E.\ E.\ Flanagan, Phys.\ Rev.\ D, {\bf 58}, 124030 (1998).

\bibitem{WilsonMathews} J.\ R.\ Wilson and G.\ J.\ Mathews, Phys.\ Rev.\ 
Lett.\ {\bf 75}, 4161 (1995); 
J.\ R.\ Wilson, G.\ J.\ Mathews and P.\ Maronetti, 
Phys.\ Rev.\ D {\bf 54}, 1317 (1996); G.\ J.\ Mathews, P.\ Maronetti, and 
J.\ R.\ Wilson, in press (gr-qc/9710140).  For the explicit identification
of an error in the Wilson-Mathews-Maronetti analysis, which presumably
produced the stellar compression, see E.\ E.\ Flanagan, Phys.\ Rev.\ D, 
submitted (gr-qc/9811132).

\bibitem{Science} B.\ A.\ Smith {\it et al.}, Science {\bf 204}, 951 (1979);
L.\ A.\ Morabito {\it et al.}, Science {\bf 204}, 972 (1979).

\bibitem{Ross} M.\ N.\ Ross and G.\ Schubert, Icarus {\bf 64}, 391 (1985).

\bibitem{Ojak} G.\ W.\ Ojakangas and D.\ J.\ Stevenson, Icarus {\bf 66}, 
341 (1986).

\bibitem{CoxGiuli} J.\ P.\ Cox and R.\ T.\ Giuli, {\it Principles of Stellar
Structure} v.1, (Gordon and Breach, Science Publishers, New York, 1968).

\bibitem{LanLif} L.\ D.\ Landau and E.\ M.\ Lifshitz, {\it The Classical 
Theory of Fields} 2nd ed., (Addison-Wesley Publishing Company, Inc., Reading, 
Massachusetts, 1962).

\bibitem{Chandra} S.\ Chandrasekhar, Astrophys. J. {\bf 158}, 45 (1969). 

\end{references}
\end{document}